\documentstyle[fleqn,11pt]{book}
\large
\begin{document}
\parindent 1.4cm
\begin{center}
{{\bf THE RAMSAUER-TOWNSEND EFFECT AND THE de BROGLIE-BOHM
QUANTUM MECHANICS}}
\end{center}
\begin{center}
{{\bf J. M. F. Bassalo$^{1}$,\ P. T. S. Alencar$^{2}$,\ A.
Nassar$^{3}$\ and\ M. Cattani$^{4}$}}
\end{center}
\begin{center}
{$^{1}$\ Funda\c{c}\~ao Minerva,\ R. Serzedelo Correa 347, 1601\
- CEP\ 66035-400,\ Bel\'em,\ Par\'a,\ Brasil}
\end{center}
\begin{center}
{E-mail:\ bassalo@amazon.com.br}
\end{center}
\begin{center}
{$^{2}$\ Universidade Federal do Par\'a\ -\ CEP\ 66075-900,\ Guam\'a,
Bel\'em,\ Par\'a,\ Brasil}
\end{center}
\begin{center}
{E-mail:\ tarso@ufpa.br}
\end{center}
\begin{center}
{$^{3}$\ Extension Program-Department of Sciences, University of California,\
Los Angeles, California 90024,\ USA}
\end{center}
\begin{center}
{E-mail:\ nassar@ucla.edu}
\end{center}
\begin{center}
{$^{4}$\ Instituto de F\'{\i}sica da Universidade de S\~ao Paulo. C. P.
66318, CEP\ 05315-970,\ S\~ao Paulo,\ SP, Brasil}
\end{center}
\begin{center}
{E-mail:\ mcattani@if.usp.br}
\end{center}
\par
ABSTRACT\ -\ In this work we study the Ramsauer-Townsend effect.
First, we use the Quantum Mechanical Formalism of
Schr\"{o}dinger. After, it is calculated with the Quantum
Mechanical Formalism of de Broglie-Bohm. In this case, we use the
Kostin equation, taking into account the energy dissipation of
the electrons scattered by sharp edged potential wells.
\newpage
\par
{\bf 1)\ INTRODUCTION}
\par
In 1921,[1] the german physicist Carl Wilhelm Ramsauer (1879-1955)
studied the scattering of low energy electrons (0.75-1.1\ eV) in inert
gases argon (A), krypton (Kr) and xenon (Xe). For A, for instance, he
observed that the effective cross-section was much smaller than that
calculated by the Kinetic Theory of Gases. In 1922[2], for higher
energies it was observed a surprising modification of the cross
section.
\par
In the same year, 1922[3], the english physicist Sir John Sealy
Edward Town\-send (1868-1957) and V. A. Bailey analyzed electronic
scattering, for electrons with energy between 0.2-0.8\ eV. Using a
different method from the adopted by Ramsauer, have found that the
maximum mean free path of the electrons occurs around 0.39\ eV. This
result, that is known as the Ramsauer-Towsend effect, was confirmed by
Ramsauer and R. Kollath in 1929[4]. This would imply that the noble
gases were transparent for a critical energy value[5,6].
\par
{\bf 2)\ THE RAMSAUER-TOWNSEND EFFECT AND THE}
\par
\ \ \ \ \ {\bf SCHR\"{O}DINGER'S EQUATION}
\par
As is well known, the linear Schr\"{o}dinger's equation is defined by
\begin{center}
{$i\ {\hbar}\ {\frac {{\partial}{\psi}(x,\ t)}{{\partial}t}}\ =\ -\
{\frac {{\hbar}^{2}}{2\ m}}\ {\frac {{\partial}^{2}\ {\psi}(x,\
t)}{{\partial}x^{2}}}\ +\ V(x,\ t)\ {\psi}(x,\ t)$\ .\ \ \ \ \ (2.1)}
\end{center}
\par
Let us consider an stationary electronic flux with incident energy E,
colliding with a potential well with height V and width L:
\begin{center}
{$V(x,\ t)\ =\ 0\ ,\ \ \ \ \ \ \ \ \ \ \ x\ {\not=}\ 0,\ L$,\ \ \ \ \
(2.2a)}
\end{center}
\begin{center}
{$V(x,\ t)\ =\ -\ V\ ,\ \ \ \ \ \ \ \ \ \ 0\ < x\ <\ L$\ ,\ \ \ \ \
(2.2b)}
\end{center}
which define the following regions:
\begin{center}
{Incidence Region (1):\ \ \ $x\ <\ 0$,\ \ \ \ \ (2.3a)}
\end{center}
\begin{center}
{Scattering Region (2):\ \ \ $0\ <\ x\ <\ L$,\ \ \ \ \ (2.3b)}
\end{center}
\begin{center}
{Transmission Region (3):\ \ \ $x\ >\ L$.\ \ \ \ \ (2.3c)}
\end{center}
\par
Since E\ $>$\ 0, the solution of the Schr\"{o}dinger's equation (2.1),
fot these three mentioned above regions is given by[7]
\begin{center}
{${\psi}_{1}(x,\ t)\ =\ (e^{i\ k_{1}\ x}\ +\ A\ e^{-\ i\ k_{1}\ x})\
e^{-\ i\ {\omega}\ t}$\ ,\ \ \ \ \ (2.4)}
\end{center}
\begin{center}
{${\psi}_{2}(x,\ t)\ =\ (C\ e^{i\ k_{2}\ x}\ +\ D\ e^{-\ i\ k_{2}\ x})\
e^{-\ i\ {\omega}\ t}$\ ,\ \ \ \ \ (2.5)}
\end{center}
\begin{center}
{${\psi}_{3}(x,\ t)\ =\ (B\ e^{i\ k_{1}\ x})\ e^{-\ i\ {\omega}\ t}$\ ,\
\ \ \ \ \ (2.6)}
\end{center}
where
\begin{center}
{$k_{1}^{2}\ =\ {\frac {2\ m\ E}{{\hbar}^{2}}}\ ,\ \ \ \ \ k_{2}^{2}\
=\ {\frac {2\ m\ (E\ +\ V)}{{\hbar}^{2}}}$\ .\ \ \ \ \ (2.7a-b)}
\end{center}
\par
So,the reflection ${\mid}\ R\ {\mid}^{2}\ =\ A\ A^{*}$ and the
transmission ${\mid}\ T\ {\mid}^{2}\ =\ B\ B^{*}$ coefficients will be
given by[8]
\begin{center}
{${\mid}\ R\ {\mid}^{2}\ =\ {\frac {{\Big {(}}\ {\frac {k_{1}^{2}\ -\
k_{2}^{2}}{2\ k_{1}\ k_{2}}}\ {\Big {)}}^{2}\ sen^{2}\ (k_{2}\ L)}{1\
+\ {\Big {(}}\ {\frac {k_{1}^{2}\ -\ k_{2}^{2}}{2\ k_{1}\ k_{2}}}{\Big
{)}}^{2}\ sen^{2}\ (k_{2}\ L)}}$\ ,\ \ \ \ \ (2.8)}
\end{center}
\begin{center}
{${\mid}\ T\ {\mid}^{2}\ =\ {\frac {1}{1\ +\ {\Big {(}}\ {\frac
{k_{1}^{2}\ -\ k_{2}^{2}}{2\ k_{1}\ k_{2}}}\ {\Big {)}}^{2}\ sen^{2}\
(k_{2}\ L)}}$\ .\ \ \ \ \ (2.9)}
\end{center}
\par
Using Eqs. (2.8)-(2.9) we will analyze a particular case assuming that[9]
\begin{center}
{$L\ =\ {\frac {{\lambda}_{2}}{2}}$\ ,\ \ \ \ \ (2.10)}
\end{center}
and considering also the de Broglie ``pilot wave'', that is $k\ =\
{\frac {p}{{\hbar}}}\ =\ {\frac {2\ {\pi}\ p}{h}}$, we get:
\begin{center}
{${\lambda}_{2}\ =\ {\frac {h}{p_{2}}}\ =\ {\frac {2\ {\pi}}{k_{2}}}\
,\ \ \ \ \ k_{2}\ L\ =\ {\pi}$\ .\ \ \ \ \ (2.11-12)}
\end{center}
\par
Substituting Eqs. (2.8)-(2.9) into Eq. (2.12), results:
\begin{center}
{${\mid}\ T\ {\mid}^{2}\ =\ 1,\ \ \ \ \ {\mid}\ R\ {\mid}^{2}\ =\ 0$\
.\ \ \ \ \ (2.13a-b)}
\end{center}
\par
These equations show that when the incident electron wavelength is two
times larger than the well width L, there is no reflection and the
transmission is complete. This is the way how Schr\"{o}dinger quantum
mechanics explains the Ramsauer-Townsend effect.
\par
{\bf 3) THE RAMSAUER-TOWNSEND EFFECT AND THE}
\par
\ \ \ \ \ {\bf de BROGLIE-BOHM FORMALISM}
\par
In this section we study the Ramsauer-Townsend effect when in the
collision process there is an energy dissipation governed by Kostion
Equation that is defined by[8]
\begin{center}
{$i\ {\hbar}\ {\frac {{\partial}{\psi}(x,\ t)}{{\partial}t}}\ =\ -\
{\frac {{\hbar}^{2}}{2\ m}}\ {\frac {{\partial}^{2}{\psi}(x,\
t)}{{\partial}x^{2}}}\ +\ [V(x,\ t)\ +\ {\frac {{\hbar}\ {\nu}}{2\ i}}\
{\ell}n\ {\frac {{\psi}(x,\ t)}{{\psi}^{*}(x,\ t)}}]\ {\psi}(x,\ t)$\
,\ \ \ \ \ (3.1)}
\end{center}
where ${\psi}(x,\ t)$, V(x,\ t) and ${\nu}$ represent, respectively, the
wavefunction, potential and dissipation constant of our system. This
Eq. (3.1) will be studied within the formalism of the de Broglie-Bohm.
\par
Putting ${\psi}(x,\ t)$ as[10]
\begin{center}
{${\psi}(x,\ t)\ =\ {\Phi}(x)\ exp\ [-\ {\frac {i\ E}{{\hbar}\ {\nu}}}\
(1\ -\ e^{-\ {\nu}\ t})]$,\ \ \ \ \ (3.2)}
\end{center}
we can write,
\begin{center}
{$i\ {\hbar}\ {\frac {{\partial}{\psi}(x,\ t)}{{\partial}t}}\ =\ E\
e^{-\ {\nu}\ t}\ {\psi}(x,\ t)$\ ,\ \ \ \ \ (3.3a)}
\end{center}
\begin{center}
{$-\ {\frac {{\hbar}^{2}}{2\ m}}\ {\frac {{\partial}^{2}{\psi}(x,\
t)}{{\partial}x^{2}}}\ =\ -\ {\frac {{\hbar}^{2}}{2\ m}}\ {\frac
{{\Phi}''(x)}{{\Phi}(x)}}\ {\psi}(x,\ t)$\ ,\ \ \ \ \ (3.3b)}
\end{center}
\begin{center}
{${\frac {{\hbar}\ {\nu}}{2\ i}}\ {\ell}n\ {\frac
{{\psi}(x,\ t)}{{\psi}^{*}(x,\ t)}}\ =\ {\frac {{\hbar}\ {\nu}}{2\ i}}\
{\ell}n\ {\frac {{\Phi}(x)}{{\Phi}^{*}(x)}}\ -\ E\ (1\ -\ e^{-\ {\nu}\
t})$\ .\ \ \ \ \ (3.3c)}
\end{center}
\par
Inserting Eqs.(3.3a-c) into Eq.(3.1) and using Eq.(2.2b) we obtain,
\begin{center}
{${\Phi}''(x)\ +\ {\Big {[}}\ q^{2}\ -\ {\frac {m\ {\nu}}{i\ {\hbar}}}\
{\ell}n\ {\frac {{\Phi}(x)}{{\Phi}^{*}(x)}}\ {\Big {]}}\ {\Phi}(x)\ =\ 0\
,\ \ \ \ \ q^{2}\ =\ {\frac {2\ m}{{\hbar}^{2}}}\ (E\ +\ V)$\ .\ \ \ \
\ (3.4a-b)}
\end{center}
\par
Now, considering ${\phi}(x,\ t)$ given by the Madelung-Bohm
transformation[8]:
\begin{center}
{${\Phi}(x)\ =\ {\phi}(x)\ e^{i\ S(x)}$\ ,\ \ \ \ \ (3.5)}
\end{center}
we see that Eq.(3.4a) becomes
\begin{center}
{${\frac {d{\Phi}(x)}{dx}}\ {\equiv}\ {\Phi}'(x)\ =\ {\phi}'(x)\ e^{i\
S(x)}\ +\ i\ {\phi}(x)\ e^{i\ S(x)}\ S'(x)$}
\end{center}
\begin{center}
{${\frac {d^{2}{\Phi}(x)}{dx^{2}}}\ {\equiv}\ {\Phi}''(x)\ =\ {\frac
{d}{dx}}\ {\frac {d{\Phi}(x)}{dx}}\ =$}
\end{center}
\begin{center}
{${\Phi}''\ =\ e^{i\ S}\ [{\phi}''\ +\ 2\ i\ {\phi}'\ S'\ -\ {\phi}(x)\
(S')^{2}\ +\ i\ {\phi}\ S'']$\ ,}
\end{center}
\begin{center}
{${\phi}''\ +\ 2\ i\ {\phi}'\ S'\ -\ {\phi}\ (S')^{2}\ +\ i\
{\phi}\ S''\ +\ [q^{2}\ -\ {\frac {2\ m\ {\nu}}{{\hbar}}}\ S]\ {\phi}\
=\ 0$\ .\ \ \ \ \ (3.6)}
\end{center}
\par
Separating real and imaginary parts of the above expression we have,
\begin{center}
{${\phi}''\ +\ (q^{2}\ -\ {\frac {2\ m\ {\nu}}{{\hbar}}}\ S)\ {\phi}\
=\ (S')^{2}\ {\phi}\ ,\ \ \ \ \ 2\ {\phi}'\ S'\ +\ {\phi}\ S''\ =\ 0$\
.\ \ \ \ \ (3.7a-b)}
\end{center}
\par
Defining,
\begin{center}
{${\rho}(x)\ =\ {\phi}^{2}(x)$\ ,\ \ \ \ \ (3.8)}
\end{center}
and integrating Eq.(3.7b) results:
\begin{center}
{${\frac {(S')'}{S'}}\ =\ -\ {\frac {2\ {\phi}'}{{\phi}}}\ \ \ {\to}\ \
\ \ \ {\int}\ {\frac {(S')'}{S'}}\ =\ -\ {\int}\ {\frac {2\
{\phi}'}{{\phi}}}\ \ \ {\to}$}
\end{center}
\begin{center}
{${\ell}n\ S'\ =\ -\ 2\ {\ell}n\ {\phi}\ +\ {\ell}n\ C\ =\ -\ {\ell}n\
{\phi}^{2}\ +\ {\ell}n\ C\ =\ {\ell}n\ {\frac {C}{{\phi}^{2}}}\ \ \
{\to}$}
\end{center}
\begin{center}
{$S'(x)\ =\ {\frac {C}{{\rho}(x)}}\ ,\ \ \ \ \ S(x)\ =\ S(0)\ +\ C\
{\int}_{o}^{x}\ {\frac {dx'}{{\rho}}}$\ .\ \ \ \ \ (3.9a-b)}
\end{center}
\par
Multiplying Eq.(3.7a) by ${\phi}'$ and using Eqs.(3.8,9a) we have,
\begin{center}
{${\phi}''\ {\phi}'\ +\ [q^{2}\ -\ (S')^{2}]\ {\phi}\ {\phi}'\ =\
{\frac {2\ m\ {\nu}}{{\hbar}}}\ S\ {\phi}\ {\phi}'\ \ \ {\to}$}
\end{center}
\begin{center}
{${\frac {d}{dx}}\ [{\frac {1}{2}}\ ({\phi}')^{2}\ +\ {\frac {1}{2}}\
q^{2}\ {\phi}^{2}\ +\ {\frac {1}{2}}\ {\frac {C^{2}}{{\phi}^{2}}}]\ =\
{\frac {2\ m\ {\nu}}{{\hbar}}}\ S\ {\phi}\ {\phi}'$\ .\ \ \ \ \
(3.10)}
\end{center}
\par
Since Eq.(3.8) can be written as,
\begin{center}
{${\rho}'\ =\ 2\ {\phi}\ {\phi}'$\ ,\ \ \ \ \ (3.11)}
\end{center}
Eq.(3.10) can be rewritten, using also Eq.(3.8), as:
\begin{center}
{$I'(x)\ =\ {\frac {2\ m\ {\nu}}{{\hbar}}}\ S(x)\ {\rho}'(x)\ ,\ \ \ \ \
I(x)\ =\ {\frac {[{\rho}(x)']^{2}}{4\ {\rho}(x)}}\ +\ q^{2}\
{\rho}(x)\ +\ {\frac {C^{2}}{{\rho}(x)}}$\ .\ \ \ \ \ (3.12a-b)}
\end{center}
\par
One can easily see that, taking Eq.(3.9a), we obtain,
\begin{center}
{${\frac {d}{dx}}\ (S\ {\rho})\ =\ S'\ {\rho}\ +\ S\ {\rho}'\ =\ S\
{\rho}'\ +\ C\ =\ S\ {\rho}'\ +\ {\frac {d}{dx}}\ (C\ x)\ \ \ {\to}$}
\end{center}
\begin{center}
{$S\ {\rho}'\ =\ {\frac {d}{dx}}\ (S\ {\rho}\ -\ C\ x)$\ .\ \ \ \ \
(3.13)}
\end{center}
\par
Substituting Eq.(3.13) into Eq.(3.12a) and considering Eqs.(3.9b,12b)
results,
\begin{center}
{${\frac {dI}{dx}}\ =\ {\frac {2\ m\ {\nu}}{{\hbar}}}\ {\frac {d}{dx}}\
(S\ {\rho}\ -\ C\ x)\ \ \ {\to}\ \ \ {\frac {d}{dx}}\  [I\ -\ {\frac
{2\ m\ {\nu}}{{\hbar}}}\ (S\ {\rho}\ -\ C\ x)]\ =\ 0\ \ \ {\to}$}
\end{center}
\begin{center}
{$I\ -\ {\frac {2\ m\ {\nu}}{{\hbar}}}\ (S\ {\rho}\ -\ C\ x)\ =\
constante\ =\ I_{o}\ \ \ {\to}$}
\end{center}
\begin{center}
{$I(x)\ =\ I_{o}\ +\ {\frac {2\ m\ {\nu}}{{\hbar}}}\ [S(x)\ {\rho}(x)\
-\ C\ x]$\ ,\ \ \ \ \ (3.14a)}
\end{center}
\begin{center}
{$I_{o}\ =\ {\frac {[{\rho}'(x)]^{2}}{4\ {\rho}(x)}}\ +\ q^{2}\
{\rho}(x)\ +\ {\frac {C^{2}}{{\rho}(x)}}\ -$}
\end{center}
\begin{center}
{$-\ {\frac {2\ m\ {\nu}}{{\hbar}}}\ {\Big {[}}\ {\rho}(x)\ {\Big {(}}\
S(0)\ +\ C\ {\int}_{o}^{x}\ {\frac {dx'}{{\rho}(x')}}\ {\Big {)}}\ -\
C\ x\ {\Big {]}}$\ .\ \ \ \ \ (3.14b)}
\end{center}
\par
Now, let us solve the differential equation (3.12b) using the
Variational Parameters technique[10] putting
\begin{center}
{${\rho}(x)\ =\ {\frac {1}{2\ q^{2}}}\ {\Big {[}}\ I(x)\ +\ {\sqrt
{I^{2}(x)\ -\ 4\ q^{2}\ C^{2}}}\ {\times}\ cos\ {\Big {(}}\ 2\ q\ [x\
-\ {\beta}(x)]\ {\Big {)}}\ {\Big {]}}$\ ,\ \ \ \ \ (3.15)}
\end{center}
where ${\beta}(x)$ is the variational unknown function. To determine
${\beta}(x)$ we derive Eq.(3.15), that is,
\begin{center}
{${\rho}'(x)\ =\ {\frac {1}{2\ q^{2}}}\ {\Bigg {(}}\ I'(x)\ +\ {\frac
{I(x)\ I'(x)\ cos\ {\Big {(}}\ 2\ q\ [x\ -\ {\beta}(x)]\ {\Big
{)}}}{{\sqrt {I^{2}(x)\ -\ 4\ q^{2}\ C^{2}}}}}\ -$}
\end{center}
\begin{center}
{$-\ {\sqrt {I^{2}(x)\ -\ 4\ q^{2}\ C^{2}}}\ sen\ {\Big {(}}\ 2\ q\ [x\
-\ {\beta}(x)]\ {\Big {)}}\ {\times}\ 2\ q\ [1\ -\ {\beta}'(x)]\ {\Bigg
{)}}$\ ,\ \ \ \ \ (3.16)}
\end{center}
where the following conditions must be obeyed:
\begin{center}
{$I'(x)\ +\ {\frac {I(x)\ I'(x)\ cos\ [2\ {\theta}(x)]}{{\sqrt
{I^{2}(x)\ -\ 4\ q^{2}\ C^{2}}}}}\ +\ {\sqrt {I^{2}(x)\ -\ 4\ q^{2}\
C^{2}}}\ {\times}$}
\end{center}
\begin{center}
{${\times}\ 2\ q\ {\beta}'(x)\ sen\ [2\ {\theta}(x)]\ =\ 0\ ,\ \ \ \ \
{\theta}(x)\ =\ q\ [x\ -\ {\beta}(x)]$\ .\ \ \ \ \ (3.17a-b)}
\end{center}
\par
This implies that Eq.(3.16) is written as,
\begin{center}
{${\rho}'(x)\ =\ -\ {\frac {{\sqrt {I^{2}(x)\ -\ 4\ q^{2}\
C^{2}}}}{q}}\ sen\ [2\ {\theta}(x)]$\ .\ \ \ \ \ (3.18)}
\end{center}
\par
>From Eqs.(3.12a) and (3.18) we get,
\begin{center}
{${\beta}'(x)\ =\ {\frac {m\ {\nu}\ S(x)}{{\hbar}\ q^{2}}}\ {\Big {(}}\
1\ +\ {\frac {I(x)\ cos\ [2\ {\theta}(x)]}{{\sqrt {I^{2}(x)\ -\ 4\
q^{2}\ C^{2}}}}}\ {\Big {)}}$\ .\ \ \ \ \ (3.19)}
\end{center}
\par
We study now the scattering of a stationary flux of particles with
energy $E$ e $k^{2}\ =\ {\frac {2\ m\ E}{{\hbar}^{2}}}$ [see Eq. (2.5a)]
by a potential well defined by Eqs.(2.2a-b).
\par
The particles flux, incident ($x\ <\ 0$) and transmitted ($x\ >\ L$),
will be given by
\begin{center}
{${\psi}_{I}(x)\ =\ e^{i\ k\ x}\ +\ A\ e^{-\ i\ k\ x}\ =\ {\phi}(x)\
e^{i\ S(x)}$\ ,\ \ \ \ \ (3.20a)}
\end{center}
\begin{center}
{${\psi}_{T}(x)\ =\ B\ e^{i\ k\ x}\ =\ {\phi}(x)\ e^{i\ S(x)}$\ .\ \ \
\ \ (3.20b)}
\end{center}
\par
Since ${\psi}$ and ${\frac {{\partial}{\psi}}{{\partial}x}}$ are
assumed to be continuos at the boundaries of the potential well, we
obtain:
\par
a)\ x\ =\ 0
\par
Using Eq.(3.20a) results:
\begin{center}
{${\psi}(x=0)\ \ \ {\to}\ \ \ 1\ +\ A\ =\ {\phi}(0)\ e^{i\ S(0)}$\ ,\ \
\ \ \ (3.21a)}
\end{center}
\begin{center}
{${\frac {{\partial}{\psi}_{I}}{{\partial}x}}{\vert}_{x\ =\ 0}\ \ \
{\to}\ \ \ 1\ -\ A\ =\ {\frac {e^{i\ S(0)}}{k}}\ [-\ i\ {\phi}'(0)\ +\
{\phi}(0)\ S'(0)]$\ .\ \ \ \ \ (3.21b)}
\end{center}
\par
Adding Eqs.(3.21a-b), we get the following expression:
\begin{center}
{$2\ k = [cos\ S(0) + i\ sen\ S(0)] {\Big {(}}\ {\phi}(0)\ [k\ +\
S'(0)] - i\ {\phi}'(0)\ {\Big {)}}$ ,}
\end{center}
that, can be divided in two parts, real and imaginary:
\begin{center}
{$2\ k\ =\ cos\ S(0)\ {\phi}(0)\ [k\ +\ S'(0)]\ +\ {\phi}'(0)\ sen\
S(0)$\ ,\ \ \ \ \ (3.22a)}
\end{center}
\begin{center}
{$0\ =\ sen\ S(0)\ {\phi}(0)\ [k\ +\ S'(0)]\ -\ {\phi}'(0)\ cos\ S(0)$\
.\ \ \ \ \ (3.22b)}
\end{center}
\par
Multiplying Eq.(3.22a) by $sin\ S(0)$ and Eq.(3.22b) by $cos\ S(0)$ and
subtracting the expressions, results:
\begin{center}
{$2\ k\ sen\ S(0)\ =\ {\phi}'(0)$\ .\ \ \ \ \ (3.23a)}
\end{center}
\par
On the other hand, multiplying Eq.(3.22a) by $cos\ S(0)$ and Eq.(3.22b)
by $sin\ S(0)$ and adding the expressions, we obtain:
\begin{center}
{$2\ k\ cos\ S(0)\ =\ {\phi}(0)\ [k\ +\ S'(0)]$\ .\ \ \ \ \ (3.23b)}
\end{center}
\par
Squaring and adding Eqs.(3.23a-b) the following expression is found:
\begin{center}
{$4\ k^{2}\ =\ [{\phi}'(0)]^{2}\ +\ {\phi}^{2}(0)\ [k\ +\ S'(0)]^{2}$\
.\ \ \ \ \ (3.23c)}
\end{center}
\par
b)\ x\ =\ L
\par
Using Eq.(3.20b), we find:
\begin{center}
{${\psi}(x=L)\ \ \ {\to}\ \ \ B\ e^{i\ k\ L}\ =\ {\phi}(L)\ e^{i\
S(L)}$\ ,\ \ \ \ \ (3.24a)}
\end{center}
\begin{center}
{${\frac {{\partial}{\psi}_{T}}{{\partial}x}}{\vert}_{x\ =\ L}\ \ \
{\to}$}
\end{center}
\begin{center}
{$B\ e^{i\ k\ L}\ =\ {\frac {1}{k}}\ e^{i\ S(L)}\ {\times}\ [-\ i\
{\phi}'(L)\ +\ S'(L)\ {\phi}(L)]$\ .\ \ \ \ \ (3.24b)}
\end{center}
\par
Taking the real and imaginary parts of Eqs. (3.24a-b) and using
Eqs. (3.9a,11):
\begin{center}
{${\phi}(L)\ =\ {\frac {1}{k}}\ [-\ i\ {\phi}'(L)\ +\ S'(L)\
{\phi}(L)]\ \ \ {\to}\ \ \ {\phi}(L)\ =\ {\frac {1}{k}}\ S'(L)\
{\phi}(L)\ \ \ {\to}$}
\end{center}
\begin{center}
{$S'(L)\ =\ k\ ,\ \ \ \ \ {\rho}(L)\ =\ {\frac {C}{k}}$\ ,\ \ \ \ \
(3.25a-b)}
\end{center}
\begin{center}
{${\phi}'(L)\ =\ 0\ ,\ \ \ \ \ {\rho}'(L)\ =\ 0$\ .\ \ \ \ \
(3.26a-b)}
\end{center}
\par
Subtracting Eqs. (3.21a-b) and taking into account
Eqs. (3.8,9a,11,23a-b), we wil find:
\begin{center}
{$2\ A\ =\ e^{i\ S(0)}\ {\Big {(}}\ {\phi}(0)\ [1\ -\ {\frac
{S'(0)}{k}}]\ +\ {\frac {i}{k}}\ {\phi}'(0)\ {\Big {)}}\ \ \ {\to}$}
\end{center}
\begin{center}
{$A\ =\ {\frac  {2\ i\ [k\ {\rho}(0)\ -\ C]\ -\ {\rho}'(0)}{2\ i\ [k\
{\rho}(0)\ +\ C]\ +\ {\rho}'(0)}}$\ .\ \ \ \ \ (3.27)}
\end{center}
\par
Taking the above expression let us calculate the refletion (${\mid}\ R\
{\mid}^{2}$) and transmission (${\mid}\ T\ {\mid}^{2}$) coefficients:
\begin{center}
{${\mid}\ R\ {\mid}^{2}\ =\ A\ A^{*}\  =\ {\frac {4\ [k\ {\rho}(0)\ -\
C]^{2}\ +\ [{\rho}'(0)]^{2}}{4\ [k\ {\rho}(0)\ +\ C]^{2}\ +\
[{\rho}'(0)]^{2}}}$\ ,\ \ \ \ \ (3.28a)}
\end{center}
\begin{center}
{${\mid}\ T\ {\mid}^{2}\ =\ 1\ -\ {\mid}\ R\ {\mid}^{2}\ =\ {\frac {4\
k\ C}{{\frac {[{\rho}'(0)]^{2}}{4\ {\rho}(0)}}\ +\ {\frac
{C^{2}}{{\rho}(0)}}\ +\ k^{2}\ {\rho}(0)\ +\ 2\ k\ C}}$\ .\ \ \ \ \
(3.28b)}
\end{center}
\par
Putting $x\ =\ 0$ into Eq. (3.14), results,
\begin{center}
{$I_{o}\ =\ {\frac {[{\rho}'(0)]^{2}}{4\ {\rho}(0)}}\ + q^{2}\ {\rho}(0)\
+\ {\frac {C^{2}}{{\rho}(0)}}\ -\ {\frac {2\ m\ {\nu}}{{\hbar}}}\
{\rho}(0)\ S(0)$\ .\ \ \ \ \ (3.29)}
\end{center}
\par
Now, substituting Eq. (3.29) into Eq. (3.28b) the transmission
coefficient is written as,
\begin{center}
{${\mid}\ T\ {\mid}^{2}\ =\ {\frac {4\ k\ C}{I_{o}\ +\ [k^{2}\ -\
q^{2}\ +\ {\frac {2\ m\ {\nu}}{{\hbar}}}\ S(0)]\ {\rho}(0)\ +\ 2\ k\
C}}$\ .\ \ \ \ \ (3.30)}
\end{center}
\par
The above expression can be written in a different form. Indeed,
consi\-dering Eqs. (3.8) and (3.25b), and using Eq. (3.24a) we can write:
\begin{center}
{${\mid}\ T\ {\mid}^{2}\ =\ B\ B^{*}\ =\ {\phi}^{2}(L)\ =\ {\rho}(L)\
=\ {\frac {C}{k}}$\ .\ \ \ \ \ (3.31)}
\end{center}
\par
To obtain the final form for ${\mid}\ T\ {\mid}^{2}$ we need to
determine the constant C. To do this, it is necessary to accomplish
some intermediate steps. Thus, taking Eqs. (3.12b,25b,26b) we have,
\begin{center}
{$I(L)\ =\ k\ C\ (1\ +\ n^{2})\ ,\ \ \ \ \ n\ =\ {\frac {q}{k}}$\ .\ \
\ \ \ (3.32a-b)}
\end{center}
\par
Starting from Eqs. (3.14a,25b,32a) we will find
\begin{center}
{$I_{o}\ =\ C\ {\Big {(}}\ k\ (1\ +\ n^{2})\ +\ {\frac {2\ m\
{\nu}}{{\hbar}\ k}}\ [k\ L\ -\ S(L)]\ {\Big {)}}$\ .\ \ \ \ \ (3.33)}
\end{center}
\par
On the other hand, from Eq. (3.14a),
\begin{center}
{$I(0)\ =\ I_{o}\ +\ {\frac {2\ m\ {\nu}}{{\hbar}}}\ S(0)\ {\rho}(0)$\
.\ \ \ \ \ (3.34a)}
\end{center}
\par
Now, using Eq. (3.23c) and Eqs. (3.8,9a,11,29,32b,34a) the function
I(0) becomes,
\begin{center}
{$I(0)\ =\ 4\ k^{2}\ -\ 2\ k\ C\ -\ k^{2}\ (1\ -\ n^{2})\ {\rho}(0)$\
.\ \ \ \ \ (3.34b)}
\end{center}
\par
>From Eqs. (3.8,9a,11,17b,18,23a-b,26b,33,34a-b) we can also verify that
\begin{center}
{$S(0)\ =\ arctg\ {\Big {(}}\ {\frac {{\rho}'(0)}{2\ [k\ {\rho}(0)\ +\
C]}}\ {\Big {)}}\ ,\ \ \ \ \ {\beta}(L)\ =\ L$,\ \ \ \ \ (3.35-36)}
\end{center}
\begin{center}
{$I(0)\ =\ C\ {\Big {(}}\ k(1\ +\ n^{2})\ +\ {\frac {2\ m\
{\nu}}{{\hbar}\ k}}\ [k\ L\ -\ S(L)]\ {\Big {)}}\ +\ {\frac {2\ m\
{\nu}}{{\hbar}}}\ S(0)\ {\rho}(0)$\ ,\ \ \ \ \ (3.37a)}
\end{center}
\begin{center}
{$C\ {\Big {(}}\ k(3\ +\ n^{2})\ +\ {\frac {2\ m\
{\nu}}{{\hbar}\ k}}\ [k\ L\ -\ S(L)]\ {\Big {)}}\ =\ 4\ k^{2}\ -$}
\end{center}
\begin{center}
{$-\ {\rho}(0)\ [k^{2}\ (1\ -\ n^{2})\ +\ {\frac {2\ m\
{\nu}}{{\hbar}}}\ S(0)]$\ .\ \ \ \ \ (3.37b)}
\end{center}
\par
Substituting Eq. (3.37) into Eq. (3.15) results
\begin{center}
{$2\ q^{2}\ {\rho}(0)\ =\ I(0)\ +\ {\sqrt {I^{2}(0)\ -\ 4\ q^{2}\ C^{2}}}\
cos\ [2\ q\ {\beta}(0)]\ =$}
\end{center}
\begin{center}
{$=\ C\ {\Big {(}}\ k(1\ +\ n^{2})\ +\ {\frac {2\ m\
{\nu}}{{\hbar}\ k}}\ [k\ L\ -\ S(L)]\ {\Big {)}}\ +\ {\frac {2\ m\
{\nu}}{{\hbar}}}\ S(0)\ {\rho}(0)\ +$}
\end{center}
\begin{center}
{$+\ {\sqrt {D_{{\nu}}}}\ cos\ [2\ q\ {\beta}(0)]$\ ,\ \ \ \ \ (3.38a)}
\end{center}
where [taking only the first order terms in ${\nu}$ and using Eq.
(3.32b)]:
\begin{center}
{$D_{{\nu}}\ =\ {\Bigg {(}}\ C^{2}\ k^{2}\ (1\ +\ n^{2})^{2}\ +\ {\frac {4\ m\
{\nu}\ C^{2}\ (1\ +\ n^{2})}{{\hbar}}}\ [k\ L\ -\ S(L)]\ {\Bigg {)}}\ +$}
\end{center}
\begin{center}
{$+\ {\frac {4\ m\ {\nu}}{{\hbar}}}\ C\ k\ (1\ +\ n^{2})\ {\rho}(0)\
S(0)\ -\ 4\ k^{2}\ n^{2}\ C^{2}$\ .\ \ \ \ \ (3.38b)}
\end{center}
\par
Analyzing the above equation, we note that its first term is given by
${\rho}(0)\ {\nu}$. As only first terms in ${\nu}$ are being taken into
account, no terms involving ${\nu}$, as a factor, in the ${\rho}(0)$
expression will be considered. This term will be represented by
${\rho}_{{\nu}}$. Thus, using Eq. (3.15), we can write:
\begin{center}
{${\rho}_{{\nu}}(0)\ =\ {\frac {1}{2\ q^{2}}}\ {\Bigg {(}}\ {\sqrt
{I^{2}(0)\ -\ 4\ q^{2}\ C^{2}}}\ cos\ [2\ q\ {\beta}(0)]\ +\ I(0)\
{\Bigg {)}}$\ .\ \ \ \ \ (3.38c)}
\end{center}
\par
Now, taking Eq. (3.37a) without ${\nu}$ factors, that is, $I(0)\
{\sim}\ C\ k\ (1\ +\ n^{2})$, inserting it in Eq. (3.38c) and also
using Eq. (3.32b) results,
\begin{center}
{${\rho}_{{\nu}}(0)\ =\ {\frac {C}{n^{2}\ k}}\ {\Big {[}}\ (n^{2}\ -\ 1)\
{\times}\ cos^{2}\ [q\ {\beta}(0)]\ +\ 1\ {\Big {]}}$\ .\ \ \ \ \
(3.39)}
\end{center}
\par
Substituting Eq. (3.39) into Eq. (3.38b), putting ${\sqrt {1\ +\ x}}\
{\sim}\ 1\ +\ x/2$ for $x\ {\ll}\ 1$\ and that $q\ >\ k\ {\to}\ n\ >\
1$, according to Eqs. (2.5a) and (3.4b), we get:
\begin{center}
{${\sqrt {D_{{\nu}}}}\ =\ C\ k\ (n^{2}\ -\ 1)\ {\Bigg {(}} 1+\ {\frac
{2\ m\ {\nu}\ (1\ +\ n^{2})}{{\hbar}\ k^{2}\ (n^{2}\ -\ 1)^{2}}}\ {\Bigg
{[}}\ k\ L\ -\ S(L)\ +$}
\end{center}
\begin{center}
{$+\ {\frac {S(0)}{n^{2}}}\ {\Big {(}}\ (n^{2}\ -\ 1)\ cos^{2}\ [q\
{\beta}(0)]\ +\ 1\ {\Big {)}}\ {\Bigg {]}}\ {\Bigg {)}}$\ .}
\end{center}
\par
Considering the Eq. (3.38a) and inserting into the above expression,
the following expression is obtained for ${\rho}(0)$,
\begin{center}
{${\rho}(0)\ =\ {\frac {C\ k}{2\ [q^{2}\ -\ {\frac {m\ {\nu}\
S(0)}{{\hbar}}}]}}\ E$\ ,\ \ \ \ \ (3.40a)}
\end{center}
where
\begin{center}
{$E\ =\ (n^{2}\ -\ 1)\ {\Bigg {(}}\ 1\ +\ {\frac {2\ m\ {\nu}\ (1\ +\
n^{2})}{{\hbar}\ k^{2}\ (n^{2}\ -\ 1)^{2}}}\ {\Big {[}}\ k\ L\ -\ S(l)\
+$}
\end{center}
\begin{center}
{$+\ {\frac {S(0)}{n^{2}}}\ {\Big {(}}\ 1\ +\ (n^{2}\ -\ 1)\ cos^{2}\ [q\
{\beta}(0)]\ {\Big {)}}\ {\Big {]}}\ {\Bigg {)}}\ cos\ [2\ q\ {\beta}(0)]\ +$}
\end{center}
\begin{center}
{$+\ (1\ +\ n^{2})\ {\Big {(}}\ 1 +\ {\frac {2\ m\ {\nu}}{{\hbar}\
k^{2}\ (1\ +\ n^{2})}}\ [k\ L\ -\ S(L)]\ {\Big {)}}$\ .\ \ \ \ \ (3.40b)}
\end{center}
\par
Now, substituting Eq. (3.40a) into Eq. (3.37b), we see that
\begin{center}
{${\frac {C}{k}}\ =\ {\frac {4}{F}}$\ ,\ \ \ \ \ (3.41a)}
\end{center}
with:
\begin{center}
{$F\ =\ 3\ +\ n^{2}\ +\ {\frac {2\ m\ {\nu}}{{\hbar}\
k^{2}}}\ [k\ L\ -\ S(L)]\ +\ {\frac {[k^{2}\ (1\ -\ n^{2})\ +\ {\frac
{2\ m\ {\nu}}{{\hbar}}}\ S(0)]}{2\ [q^{2}\ -\ {\frac {m\ {\nu}\
S(0)}{{\hbar}}}]}}\ E$\ .\ \ \ \ \ (3.41b)}
\end{center}
\par
In this way, the transmission coefficient [see Eqs. (2.2a-b)] will
written in terms of the Eqs. (3.31,41a), that is,
\begin{center}
{${\mid}\ T\ {\mid}^{2}\ =\ {\frac {4}{F}}$\ .\ \ \ \ \ (3.42)}
\end{center}
\par
Finally, the Ramsauer-Townsend effect will be studied considering the
above expression for dissipative regions having small ${\nu}$ values.
First, let us use the Taylor expansion of ${\beta}(x)$ and S(x) for $x\
{\sim}\ L$:
\begin{center}
{${\beta}(x)\ =\ {\beta}(L)\ +\ {\beta}'(L)\ (x\ -\ L)\ +\
{\beta}''(L)\ {\frac {(x\ -\ L)^{2}}{2!}}\ +$\ ... ,\ \ \ \ \
(3.43a)}
\end{center}
\begin{center}
{${\beta}(0)\ =\ L\ -\ {\beta}'(L)\ L\ +\ {\beta}''(L)\ {\frac
{L^{2}}{2!}}\ +$\ ... ,\ \ \ \ \ (3.43b)}
\end{center}
\begin{center}
{$S(x)\ =\ S(L)\ +\ S'(L)\ (x\ -\ L)\ +\ S''(L)\ {\frac {(x\ -\
L)^{2}}{2!}}\ +$\ ... ,\ \ \ \ \ (3.44a)}
\end{center}
\begin{center}
{$S(0)\ =\ S(L)\ -\ S'(L)\ L\ +\ S''(L)\ {\frac {L^{2}}{2!}}\ +$\ ...
.\ \ \ \ \ (3.44b)}
\end{center}
\par
>From Eqs. (3.17b) and (3.43b), results,
\begin{center}
{$cos\ [q\ {\beta}(0)]\ {\sim}\ 1\ \ \ {\to}$}
\end{center}
\begin{center}
{${\frac {S(0)}{n^{2}}}\ {\Big {(}} 1\ +\ (n^{2}\ -\ 1)\ cos^{2}\ [q\
{\beta}(0)]\ {\Big {)}}\ =\ {\frac {S(0)}{n^{2}}}\ (1\ +\ n^{2}\ -\ 1)\
=\ S(0)$ .}
\end{center}
\par
Substituting the above relations into Eqs. (3.40b), the function E, in
the first order ${\nu}$, will be given by:
\begin{center}
{$E_{{\nu}}\ =\ (n^{2}\ -\ 1)\ {\Big {(}}\ 1\ +\ {\frac {2\ m\ {\nu}\
(1\ +\ n^{2})}{{\hbar}\ k^{2}\ (n^{2}\ -\ 1)^{2}}}\ [\ k\ L\ -\ S(L)\
+\ S(0)\ ]\ {\Big {)}}\ {\times}$}
\end{center}
\begin{center}
{${\times}\ cos\ [2\ q\ {\beta}(0)]\ +\ (1\ +\ n^{2})\
{\Big {(}}\ 1 +\ {\frac {2\ m\ {\nu}}{{\hbar}\ k^{2}\ (1\ +\ n^{2})}}\
[k\ L\ -\ S(L)]\ {\Big {)}}$\ .\ \ \ \ \ (3.45)}
\end{center}
\par
Now, taking into account only first order ${\nu}$ terms into Eq.
(3.32b), remembering that $(1\ +\ x)^{m}\ {\sim}\ 1\ +\ m\ x$, for $x\
{\ll}\ 1$], we find,
\begin{center}
{${\frac {k^{2}\ (1\ -\ n^{2})\ +\ {\frac {2\ m\ {\nu}\
S(0)}{{\hbar}}}}{2\ q^{2}\ [1\ -\ {\frac {m\ {\nu}\ S(0)}{{\hbar}\
q^{2}}}]}}\ =\ {\frac {(1\ -\ n^{2})}{2\ n^{2}}}\ {\Big {[}}\ 1\ +\
{\frac {m\ {\nu}\ S(0)}{{\hbar}\ q^{2}}}\ {\frac {n^{2}\ +\ 1}{1\ -\
n^{2}}}\ {\Big {]}}$\ .}
\end{center}
\par
Putting the above relation into Eq. (3.41b) and using also Eqs.
(3.42,45), the function F in the first order ${\nu}$ approximation
becomes:
\begin{center}
{$F_{{\nu}}\ =\ 3\ +\ n^{2}\ +\ {\frac {2\ m\ {\nu}}{{\hbar}\
k^{2}}}\ [k\ L\ -\ S(L)]\ +$}
\end{center}
\begin{center}
{$+\ {\frac {(1\ -\ n^{2})}{2\ n^{2}}}\ {\Big {[}}\ 1\ +\ {\frac
{m\ {\nu}\ S(0)}{{\hbar}\ q^{2}}}\ {\Big {(}}\ {\frac {n^{2}\ +\ 1}{1\
-\ n^{2}}}\ {\Big {)}}\ {\Big {]}}\ E_{{\nu}}$\ ,\ \ \ \ \ (3.46)}
\end{center}
and, consequently,
\begin{center}
{${\mid}\ T\ {\mid}^{2}\ =\ {\frac {4}{F_{{\nu}}}}$\ .\ \ \ \ \ (3.47)}
\end{center}
\par
Eqs. (3.45-46) show that the transmission coefficient $T^{2}$ in a
dissipative potential well [see Eq. (3.47)] depends on S(0) and S(L).
To calculate these functions we take Eq. (3.19) and Eqs. (3.25a,32a-b),
remembering that $cos\ [2\ {\theta}(L)]\ {\sim}\ 1$ and that $n\ >\ 1$,
obtaining:
\begin{center}
{${\beta}'(L)\ =\ {\frac {2\ m\ {\nu}\ S(L)}{{\hbar}\ (q^{2}\ -\
k^{2})}}\ ,\ \ \ \ \ {\beta}''(L)\ =\ {\frac {2\ m\ {\nu}\ k}{{\hbar}\
(q^{2}\ -\ k^{2})}}$\ .\ \ \ \ \ (3.48a-b)}
\end{center}
\par
Substituting the above Eqs. (3.48a-b) into Eq. (3.43b), ${\beta}(0)$
will given by
\begin{center}
{${\beta}(0)\ {\sim}\ L\ {\Big {(}}\ 1\ +\ {\frac {2\ m\ {\nu}}{{\hbar}\
(q^{2}\ -\ k^{2})}}\ {\Big {[}}\ {\frac {k\ L}{2}}\ -\ S(L)\ {\Big {]}}\
{\Big {)}}$\ .\ \ \ \ \ (3.49)}
\end{center}
\par
Since $k^{2}\ =\ {\frac {2\ m\ E}{{\hbar}^{2}}}$, Eqs. (3.4b,9a,26b)
can be written as follows,
\begin{center}
{$q^{2}\ -\ k^{2}\ =\ {\frac {2\ m}{{\hbar}^{2}}}\ (E\ +\ V)\ -\ {\frac
{2\ m\ E}{{\hbar}^{2}}}\ =\ {\frac {2\ m\ V}{{\hbar}^{2}}}\ ,\ \ \ \ \
S''(L)\ =\ 0$\ .\ \ \ \ \ (3.50a-b)}
\end{center}
\par
This permit us to write Eqs. (3.25a,44b,49,50b) as,
\begin{center}
{$S(0)\ {\sim}\ S(L)\ -\ k\ L$\ ,\ \ \ \ \ (3.51a)}
\end{center}
\begin{center}
{${\beta}(0)\ {\sim}\ L\ {\Big {(}}\ 1\ -\ {\frac {{\nu}\ {\hbar}}{V}}\
{\Big {[}}\ {\frac {k\ L}{2}}\ +\ S(0)\ {\Big {]}}\ {\Big {)}}$\ ,\ \ \ \ \
(3.51b)}
\end{center}
\begin{center}
{${\beta}^{2}(0)\ {\sim}\ L^{2}\ {\Big {(}}\ 1\ -\ {\frac {2\ {\nu}\
{\hbar}}{V}}\ {\Big {[}}\ {\frac {k\ L}{2}}\ +\ S(0)\ {\Big {]}}\ {\Big
{)}}$\ .\ \ \ \ \ (3.51c)}
\end{center}
\par
Now, substituting Eq. (3.51a) into Eqs. (3.45-47) and remembering that
only first order ${\nu}$ terms are being considered, we obtain
\begin{center}
{$E_{{\nu}}[S(0)]\ =\ (n^{2}\ -\ 1)\ cos\ [2\ q\ {\beta}(0)]\ +\ 1\ +\
n^{2}\ -\ {\frac {2\ m\ {\nu}\ S(0)}{{\hbar}\ k^{2}}}$\ ,\ \ \ \ \
(3.52a)}
\end{center}
\begin{center}
{$F_{{\nu}}[S(0)]\ =\ 4\ {\Bigg {(}}\ 1\ +\ sen^{2}\ [q\ {\beta}(0)]\
{\Big {(}}\ {\frac {1\ -\ n^{2}}{2\ n}}\ {\Big {)}}^{2}\ {\times}$}
\end{center}
\begin{center}
{${\times}\ {\Big {[}}\ 1\ -\ {\frac {m\ {\nu}\ S(0)}{{\hbar}\ q^{2}}}\
{\Big {(}}\ {\frac {n^{2}\ +\ 1}{n^{2}\ -\ 1}}\ {\Big {)}}\ {\Big {]}}\
{\Bigg {)}}$\ ,\ \ \ \ \ (3.52b)}
\end{center}
\begin{center}
{${\mid}\ T\ {\mid}^{-\ 2}\ =\ 1\ +\ sen^{2}\ [q\ {\beta}(0)]\ {\Big
{(}}\ {\frac {1\ -\ n^{2}}{2\ n}}\ {\Big {)}}^{2}\ {\times}$}
\end{center}
\begin{center}
{${\times}\ {\Big {[}}\ 1\ -\ {\frac {m\ {\nu}\ S(0)}{{\hbar}\ q^{2}}}\
{\Big {(}}\ {\frac {n^{2}\ +\ 1}{n^{2}\ -\ 1}}\ {\Big {)}}\ {\Big
{]}}$\ .\ \ \ \ \ (3.52c)}
\end{center}
\par
Note that Eq. (2.8) is obtained, putting ${\nu}\ =\ 0$ into Eq. (3.52c)
and using Eqs. (2.6a-b) and (3.4b,49).
\par
After these tedious calculations we are almost in conditions to
write the final expression to explain the Ramsauer-Townsend
effect. Indeed, as was shown in Secction 2, this effect is
characterized [see Eq. (2.12)] by:
\begin{center}
{$q\ L\ =\ {\pi}$\ .\ \ \ \ \ (3.53a)}
\end{center}
\par
In this way, with Eq. (3.53a), Eq. (3.51b) can be written as,
\begin{center}
{$q\ {\beta}(0)\ {\sim}\ {\pi}\ \ \ {\to}\ \ \ 2\ q\ {\beta}(0)\
{\sim}\ 2\ {\pi}$\ .\ \ \ \ \ (3.53b-c)}
\end{center}
\par
The validity of Eqs. (3.17b,18,53b-c) permit us to put,
\begin{center}
{${\rho}'(0)\ =\ 0\ ,\ \ \ \ \ sen\ [q\ {\beta}(0)]\ =\ 0$\ .\ \ \ \
\ (3.54a-b)}
\end{center}
\par
>From Eqs. (3.11,23a,54a), results
\begin{center}
{$2\ k\ sen\ S(0)\ =\ {\phi}'(0)\ =\ {\frac {{\rho}'(0)}{2\
{\phi}(0)}}\ =\ 0\ \ \ {\to}\ \ \ S(0)\ =\ 0$\ .\ \ \ \ \ (3.55)}
\end{center}
\par
Finally, Eqs. (3.52c,54b) show that the transmission coefficient is
given by:
\begin{center}
{${\mid}\ T\ {\mid}^{2}\ =\ 1$\ ,\ \ \ \ \ (3.56)}
\end{center}
in agreement with Eq. (2.13a), which characterizes the
Ramsauer-Townsend effect.
\vspace{0.2cm}
\par
{\bf NOTES AND REFERENCES}
\begin{enumerate}
\item RAMSAUER, C. W. 1921. {\it Annalen der Physik 64}, p. 513.
\item RAMSAUER, C. W. 1921. {\it Annalen der Physik 66}, p. 545.
\item TOWNSEND, J. S. E. and BAILEY, V. A. 1922. {\it Philosophi\-cal
Magazine 43; 44}, p. 593; 1033.
\item RAMSAUER, C. W. und KOLLATH, R. 1929. {\it Annalen der Physik 3}, p.
536.
\item Stephen G. Kukolich, in 1968, {\it American Journal of Physics
36}, p. 701), has shown this effect for the Xenonian (Xe).
\item For more details about this effect, see MOTT, N. F. and
MASSEY, H. S. W. 1971. {\bf The Theory of Atomic Collisions}, Clarendon
Press, Oxford; BRODE, R. B. 1933. {\it Reviews of Modern Physics 5} (p.
257.
\item The scattering of particles by potential wells is studied in many
textbooks. See, for instance:
\par
.\ BOHM, D. 1951. {\bf Quantum Theory}. Prentice-Hall, Inc.
\par
.\ DAVYDOV, A. S. 1968. {\bf Quantum Mechanics}. Addison-Wesley Pu\-blishing
Company, Inc.
\par
.\ MESSIAH, A. 1961. {\bf Quantum Mechanics I}. North-Hol\-land Publications
Company.
\par
.\ LEITE LOPES, J. 1992. {\bf A Estrutura Qu\^antica da Mat\'e\-ria}.
Editora UFRJ/ERCA Editora e Gr\'afica.
\par
.\ SCHIFF, L. I. 1955. {\bf Quantum Mechanics}. MacGraw-Hill Book Company,
Inc.
\par
.\ SPROULL, R. L. and PHILLIPS, W. A. 1980. {\bf Modern Physics}. John
Wiley and Sons.
\item BASSALO, ALENCAR, P. T. S., CATTANI, M. S. D. e NASSAR, A. B.
2003. {\bf T\'opicos da Mec\^anica Qu\^antica de de Broglie-Bohmo}. EDUFPA.
\item SPROULL and PHILLIPS, op. cit.
\item NASSAR, A. B. 1998. {\bf Effect of dissipation on scaterring and
tunneling through sharp-edged potential barriers} (DFUFPA, preprint).
\end{enumerate}
\end{document}